# Goldstone Apple Valley Radio Telescope Monitoring Flux Density of Jupiter's Synchrotron Radiation during the Juno Mission

T. Velusamy[1], V. Adumitroaie[1], J. Arballo[1], S. M. Levin[1], P. A. Ries[1],
R. Dorcey[2], N. Kreuser-Jenkins[2], J. Leflang[2]
D. Jauncey[3], S. Horiuchi[4]

[1]Jet Propulsion Laboratory, California Institute of Technology, Pasadena, CA 91109, USA
[2]Lewis Center for Educational Research, 17500 Mana Road, Apple Valley, CA 92307
[3]CSIRO/ATNF, PO Box 76, Epping NSW 1710, Australia
Research School of Astronomy and Astrophysics, ANU, Canberra 2611, Australia
[4]CSIRO Astronomy and Space Science, CDCC, PO Box 1035, Tuggeranong, ACT 2901, Australia

Corresponding Author: Thangasamy Velusamy
Email: thangasamy.velusamy@jpl.nasa.gov

Abstract

Goldstone Apple Valley Radio Telescope (GAVRT) is a science education partnership among NASA, the Jet Propulsion Laboratory (JPL), and the Lewis Center for Educational Research (LCER), offering unique opportunities for K -12 students and their teachers. As part of a long-term Jupiter synchrotron radiation (JSR) flux density monitoring program, LCER has been carrying out Jupiter observations with some student participation. In this paper we present the results of processed data sets observed between March 6, 2015 and April 6 2018. The data are divided into 5 epochs, grouped by time and by Jupiter's ephemeris parameters as observed from Earth. We derive JSR beaming curves at different epochs and Earth declinations. We present a comparison of the observed beaming curves with those derived from most recent models for the radiation belts. Our results show an increasing trend of the JSR flux density which seem consistent with the models for the magnetospheric solar wind interactions.

**Key Words:** planets and satellites: individual (Jupiter) – radio continuum: planetary systems – radiation mechanisms: non-thermal

## 1 Introduction

The NASA-JPL Jupiter flux density monitoring program has now entered its 50$^{th}$ year (Klein et al., 1972). Jupiter monitoring has been using the Goldstone Apple Valley Radio Telescope (GAVRT), since GAVRT's inception almost 25 years ago (c.f. Klein et al., 1989). GAVRT is a science education Partnership between the Jet Propulsion Laboratory, JPL, and the Lewis Center for Educational Research, LCER, in Apple Valley, California. It offers unique educational opportunities for students in grades K – 12 throughout the US and internationally, to learn about science by participating in real scientific research using a 34 m NASA radio telescope (see Roller & Klein, 2003 and Jauncey et al., 2017). The flux density measurements of Jupiter's synchrotron radio emission presented in this paper were made through the GAVRT program at 2280 MHz (13 cm wavelength) using the DSS13, one of NASA's Deep Space Network (DSN) telescopes. Here we present the results of the observations made between 2015 and 2018, laying the framework for future flux density monitoring and the prospects for further scientific research. We demonstrate that in the Juno era, the ground-based monitoring of Jupiter's synchrotron emission is relevant to the on-orbit observations by the Microwave Radiometer (MWR, Janssen et al. 2017) and to other Juno *in situ* measurements of the magnetosphere by MAG, JADE, JEDI and WAVES (e.g. Nichols et al. 2020).

The rationale for studying time variability in Jupiter's synchrotron radiation (JSR) is summarized in the review by de Pater & Klein (1989) and in the recent simulations of JSR time variations incorporating the effects of solar wind (dynamic) ram pressure (Han et al. 2018). The non-thermal component of Jupiter's flux densities at decimeter wavelengths measures the synchrotron radiation emitted by relativistic electrons, with energies up to tens of MeV, trapped in the Jovian van Allen belt, as they spiral up and down magnetic field lines and mirror at high field values (e.g. Santos-Costa et al. 2017; de Pater, 1990 ; de Pater & Klein, 1989; Berge & Gulkis, 1976). Because the synchrotron emission from relativistic electrons is highly directional (radiation is emitted within a narrow angle beam < 1°, centered in the particle's direction of motion), the spatial and temporal structures of Jupiter's decimeter non-thermal radio emission are determined by a combination of electron pitch angle distribution and magnetic field direction. Morphologically, the two-dimensional radio brightness maps are characterized by strong equatorial emission with two bright lobes on either side of the disk.

The total flux density and the spatial structure as observed from Earth change with Jupiter's rotation and are functions of the central meridian longitude (CML) and latitude as seen from Earth (e.g. de Pater, 1980). Modulation of the total non-thermal flux density caused by Jupiter's rotation (referred to as "the beaming curve" expressing the flux density as function of CML or the System III longitude) is characterized by two maxima and two minima which are governed by the dominant dipole magnetic environments within the radiation belts and the viewing geometry towards the Earth. The maxima occur for the longitudes when the Earth-based observer is roughly on the magnetic equator, while the minima occur at longitudes when one of the magnetic poles is pointing maximally (~10°) towards the Earth (see Fig. 58 in de Pater & Klein 1989;). The temporal variability of Jupiter's synchrotron radiation and its brightness distribution have been investigated by developing models of the radiation belts. A comparison of the results from a computational model of the synchrotron radiation with ground-based radio observations helps to improve or confirm the assumed processes and the physical conditions close to Jupiter affecting synchrotron emission, including the electron energy spectra, pitch angle distributions, and the magnetic environment (cf. Santos-Costa & Bourdarie, 2001; de Pater 1981; Garrett et al. 2005).

The observed variability, after correcting for effects of viewing geometry, is interpreted as the result of intrinsic changes in energetic electron population induced by some physical process. In almost all theories of Jovian radiation belts the solar ultraviolet (UV)/extreme ultraviolet (EUV) heating of Jupiter's upper atmosphere is regarded as driving neutral wind perturbations which leads to radial diffusion of energetic electrons (Brice & McDonough, 1973). Many attempts to understand temporal variations of JSR have been made by constructing various radial diffusion models (de Pater & Goertz, 1990, 1994; Miyoshi et al., 1999; Santos-Costa et al., 2008; Tsuchiyaet al., 2011). The long-term changes in the synchrotron radiation were correlated with a fluctuation in the dynamic solar wind pressure with an approximate two-year lag (Bolton et al., 1989). For variations measured over multiple days or weeks, different processes may be at play, such as solar UV/EUV flux enhancements driving particle transport (Miyoshi et al., 1999). The short-term variations are explained in terms of an enhanced diffusion coefficient in their model, taking account of the observed enhanced solar UV/EUV flux (Miyoshi et al. 1999; Tsuchiya et al. 2011). Several short-term variations not related to the solar UV/EUV activity have also been reported (Klein, 1976); in particular, impacts by comets or asteroids did affect the JSR (e.g., Shoemaker-Levy 9: de Pater et al., 1995, and the 2009 impact: Santos-Costa et al., 2011). A short-term dip (at 8% level) in the total intensity at 13 cm was observed from mid-January to mid-February, 2001 (Klein et al. 2001), but to date, no explanation has been provided for this unusual period of variability.

Time variability of the radio spectrum has been observed, but the measurements have not been sufficient to provide complete understanding of the particle source cycle, energy losses and magnetic field environment. Ground-based microwave observations have been used to adjust models which were originally based on *in situ* (Pioneer and Voyager) measurements of the electron distribution (Garrett et al. 2005). For example, Adumitroaie et al. (2019a, 2019b) and Santos-Costa et al. (2017) have incorporated Juno mission results to test and improve the computational model of synchrotron emission relying on assumed electron distributions and a Jovian magnetic field model - updated from the VIP4 framework (Connerney *et al.* 1998) to the latest incarnation, JRM09 (Connerney *et al.* 2018). Most recently, Han et al. (2018) have used long-term monitoring of EUV from Io plasma torus by the spectrometer EXCEED on board the HISAKI satellite (Yamazaki et al. 2014) and the fluctuating dawn-to-dusk electric field data to account for long term time variability. Their simulations show good correlation for the periods between 1971 and 2005; however, they find difficulties accounting for both short-term and long-term variations using the same model.

Correlation studies with particles injection and their transport, solar-wind conditions, solar radio fluxes, using physical electron radiation-belt models need availability of continuous monitoring of Jupiter's synchrotron emission. At present, most observational data available in the literature are limited to periods prior to 2005. The need for data after 2016 is particularly important, in view of space-based data from MWR on Juno (Bagenal et al., 2017; Janssen et al. 2017). In this paper we present the most recent observations, spanning March 2015 to April 2018. The observations and flux density calibrations are described in section 2. Our results and discussion of the non-thermal synchrotron radiation are presented in section 3. The long-term and plausible short-term variability in this GAVRT data are discussed. Our focus in this paper is to provide an overview of the GAVRT data and lay out a framework for citizen science monitoring of JSR.

## 2   Observations

The observations were made between March 6, 2015 and April 6, 2018 using the GAVRT 34-meter telescope and the S-band receiver operating at 2.280 GHz (13 cm wavelength) over a bandwidth of 40 MHz. The observations are summarized in Table 1. Though the GAVRT program is intended to have robust school teacher/student participation, unfortunately, it has not been always possible to involve schools. Over most of our observations, Jupiter was up from evening to the early morning hours and it has not been feasible to schedule observations during nominal school hours. In 2016, on May 24 students from Bransford High School, CT (Vicki Climie) participated in the S-band (13cm wavelength) observations of Jupiter and on May 31 some X-band observations (8.4 GHz, not presented here) were made as part of the Gaum Teacher Training program. In 2018 one of the US east coast schools, Newport Middle School - Newport, PA (led by Ms. Peggy Olson) was able to participate for the observations on January 23. However, the GAVRT team members (Nancy Kreuser-Jenkins & Kelli Cole) at LCER continued taking data monitoring JSR. In 2019, Kalee Tock, who is teaching a Stanford Online High School's Astronomy Research Seminar class, has involved four high school students in the data analysis using some of the data sets observed in 2018. This work resulted in a report and a presentation at 2019 Citizen Science Conference: "An Automated Approach to Modeling Jupiter's Synchrotron Radiation from Radio Telescope Observations" by Peyton Robertson, Connor Espenshade, Jay Sarva, Owen Dugan, & Kalèe Tock (Stanford Online High School).

The duration of Jupiter observing sessions varied from day to day and in general, with exception of 2015 DOY 65, only a few flux density measurements were made on any given day, sampling just a few data points for the System III longitude coverage. To cover the full 0°- 360° longitude we would require 9.9 hr (Jupiter's rotation period) of continuous observation which is not always feasible. Therefore, to study the variation of synchrotron emission with Jupiter's rotation, it was necessary to combine measurements made over many days during the duration of each epoch (see Table 1). However, for the epoch 2015_065 we had a long stretch of data over 11 hrs, providing adequate longitude coverage from a single day's observation. Table 1 describes the data included in each epoch. Typically, each observation consisted of a pair of scans across either Jupiter or a calibration source, along declination (dec) and along cross-declination (xdec). The scans were typically 2 min duration at the rate of 1.2 degree/min and sampling the radiometer output (antenna temperature) at 1 s interval. The scans included sufficient off-source baseline data to fit any drifts in receiver power level and/or off source emission for background subtraction. Figure 1 shows an example of scans across a calibrator and Jupiter.

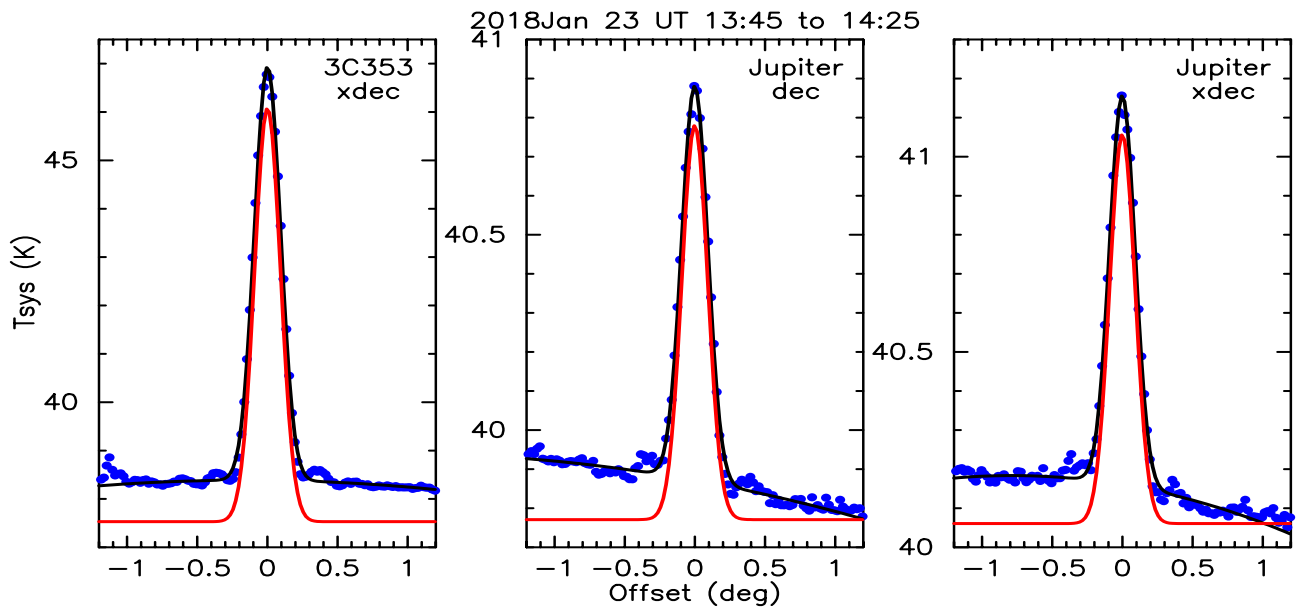

**Figure 1.** Examples of scan data and Gauss fit analysis. The data points are shown in blue. The black solid line shows the fit to the scan data with a Gaussian profile to the scan peak and a quadratic baseline. The Gaussian source profile alone is shown in red.

The peak antenna temperature and beam width were measured by simultaneously fitting a quadratic baseline and a Gaussian profile as demonstrated in Figure 1. The Gaussian fitting gives robust source antenna temperature, $T_A$ (K) with typical 1-$\sigma$ rms noise ~ 0.02 K, even in the extreme cases < 0.1 K and good fit to the baseline. The DSS-13 S-band receiver is a total power radiometer and is susceptible to temperature and atmosphere changes. These fluctuations can cause some baseline drift in the scan data (Fig. 1). However, a quadratic baseline fit works well to remove such drifts.

**Table 1:** GAVRT 13 cm (2.280 GHz) Jupiter Observations: 2015-2018 Flux density Calibration

| EPOCH | | Source | Flux density (Jy) | No. of scans | Elevation (deg) | Conversion factor* $F_{Jy}= S(Jy)/T_A(K)$ |
| --- | --- | --- | --- | --- | --- | --- |
| YEAR | DOY | | | | | |
| 2015 | 65 Mar 06 | 3C123 | 31.73 | 16 | 18 - 77 | 5.08±0.07 |
| | | 3C286 | 11.63 | 14 | 50 - 68 | 5.07±0.05 |
| | | Jupiter | -- | 222 | 19 - 72 | 5.07±0.04** |
| 2016 | 112 – 146 Apr 21 – May 25 | 3C123 | 31.73 | 9 | 46 - 70 | 4.83±0.06 |
| | | 3C218 | 27.03 | 50 | 20 - 43 | 4.78±0.11 |
| | | 3C286 | 11.63 | 12 | 74 - 85 | 4.84±0.14 |
| | | Jupiter | -- | 139 | 24 - 63 | 4.79±0.10** |
| 2017 | 3 – 69 Jan 03 – Mar 10 | 3C353 | 40.09 | 159 | 24 - 54 | 4.72±0.05 |
| | | Jupiter | -- | 334 | 22 - 47 | 4.72±0.05** |
| 2018[++] | 8 – 39 Jan 08 – Feb 8 | 3C218 | 27.03 | 5 | 17 - 20 | 4.72±0.02 |
| | | 3C286 | 11.63 | 63 | 23 - 83 | 4.70±0.06 |
| | | 3C353 | 40.09 | 331 | 23 - 56 | 4.70±0.04 |
| | 39 – 96 Feb 28 – Apr 6 | Jupiter | -- | 1024 | 23 - 54 | 4.70±0.04** |

* We define flux density conversion factor $F_{Jy}$ as $S(Jy)/T_A(K)$ where, S is source flux density and $T_A$ is the scan peak antenna temperature.
** conversion factor, $F_{Jy}$ used for calibrating Jupiter data
[++] The data were analyzed as two epochs (i) DOY 8 -39; (ii) DOY 59 – 96

3C123, 3C218, 3C286 and 3C353 were observed as calibration sources as summarized in Table 1. After 2017, 3C353 was observed on most days as it was found to be a good calibration source for Jupiter observations. Because both 3C218 and 3C353 are extended (>3 arcmin) complex sources, compared to the antenna beam size (13.6 arcmin), we first measured its flux density within the GAVRT antenna beam using 3C286 as primary flux calibrator. From the observed antenna temperatures for 3C218, 3C353 and 3C286, all available within the observations reported here, we measured their flux densities, adopting the flux density for 3C286 as S(2280MHz) = 11.63 Jy from Perley & Butler (2017), hereafter, referred to as PB2017. To be consistent with the other two calibrators, we also determined the flux density for 3C123 from our data using 3C286 as primary flux calibrator. We measure a flux density of 31.73Jy which is about 1% less than the value derived from the coefficients given in Table 6 in PB2017. It may be noted such difference may be consistent with the 3C123 nucleus being less strong than prior to 1975 (c.f. Perley & Butler, 2017). Furthermore, our measured flux densities of the calibration sources are all within 1% of the values calculated using the coefficients given by PB2017. Thus, all JSR flux measurements presented here correspond to PB2017 flux density scale.

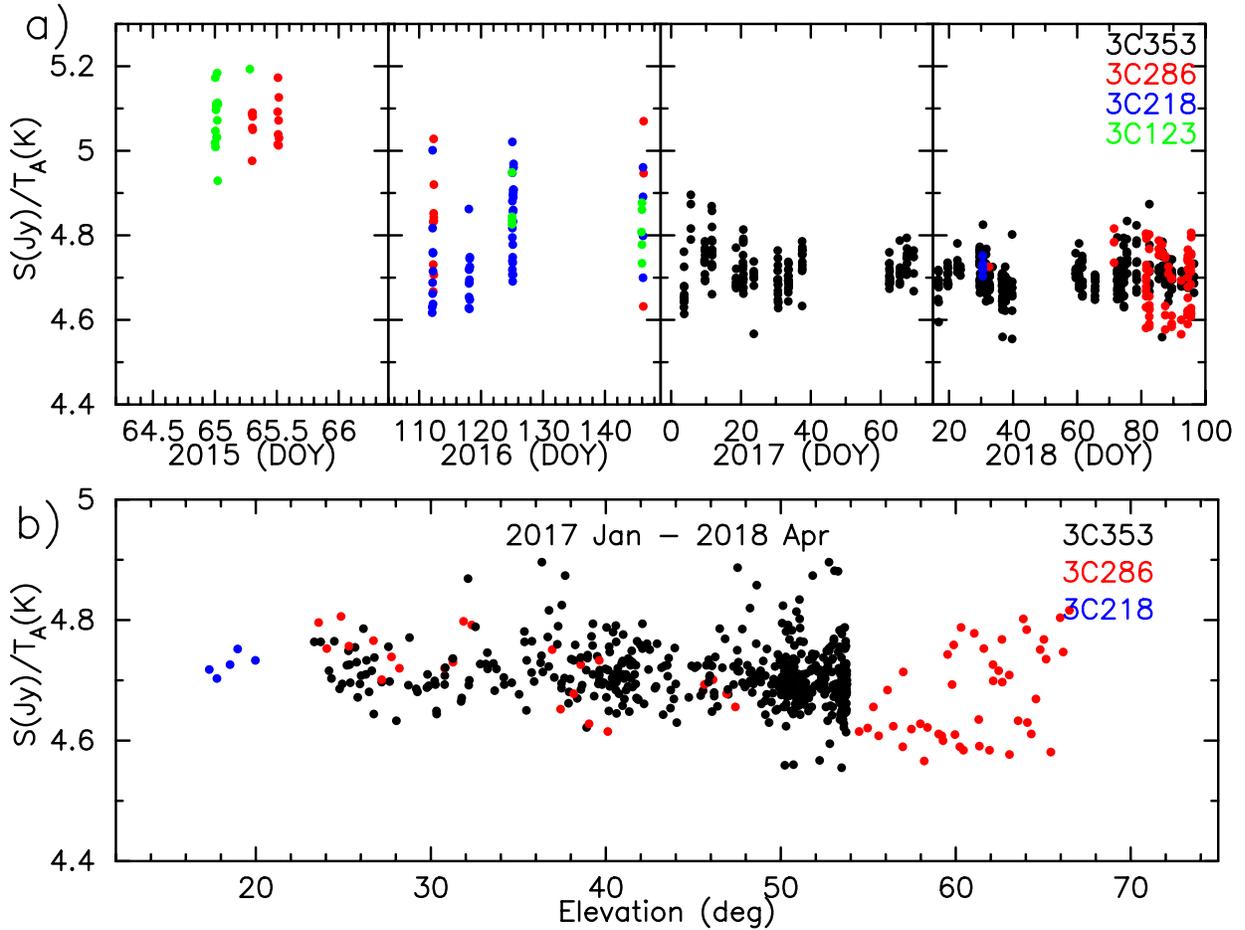

**Figure 2**. The antenna system calibration. $S(Jy)/T_A(K)$ (antenna temperature to flux density conversion factor, $F_{Jy}$) plotted as functions of time (upper panel) and elevation (lower panel) for the full period 2015 to 2018. Data points for different calibrator sources are distinguished by the colors.

From each calibration source (CAL) scan observation we derive the antenna temperature to flux density conversion factor, defined as $F_{Jy} = S(Jy)/T_A(K)$. In Figure 2 the measured flux conversion factor for each scan observed within each epoch period are plotted as a function of time (a) and source elevation (b). The conversion factor $S(Jy)/T_A(K)$, shown in Fig. 2, derived from all calibration sources seem to be very consistent within a range of 4.6 to 5.2 decreasing slightly over time from 2015 to 2018. The extreme values showing the largest scatter are the result of poor pointing. Remarkably, the gain derived from 3C353 is very stable with time between 2017 and 2018 as well as over a wide range of elevation with an average value of 4.7. The antenna calibration is summarized in Table 1 wherein are listed the flux density conversion factor ($F_{Jy}$) for each source, averaged over the period of each epoch. For a comparison, the range of elevations for each source observed at each epoch are also listed in Table 1. The flux conversion factor, $F_{Jy}$ used for calibrating the Jupiter $T_A$ into flux density (Jy) for each epoch was then obtained by averaging scan values for the calibration sources overlapping Jupiter's elevations as listed in Table 2. The near constant scan to scan value of $F_{Jy}$ suggest the recorded system antenna temperatures (Fig. 1) are well calibrated as degrees Kelvin (K) by the use of "mini cal" available in the DSS13 receiver system used for our observations (Stelzried et al. 2008). Furthermore, the mean values of the $S(Jy)/T_A(K) = 5.1$ to 4.7 derived for our multi-epoch data indicate an aperture efficiency of 58% to 64% which is consistent with the atmosphere free efficiency measurements of 71.5 ± 1.8 % for the 34-m at F3 focal point (c.f. Imbriale, 2002).

Figure 3a shows examples of the measured $T_A$ (K) with Gaussian fitting and 1-σ rms noise observed on Jupiter on 2 days in 2018. The scatter in the scan to scan measurements for a given longitude is dominated mostly by the pointing inaccuracies. As an example of the calibrated scan flux density, in Figure 3b are shown the Jupiter flux densities observed on January 29-30, 2018. The error bars correspond to the 1-σ uncertainty for the Gauss fit peak temperatures and the $T_A$ to flux conversion. We did not find any systematic differences between the dec and xdec scan antenna temperatures (or flux densities). Therefore, in all our analysis we do not distinguish between the scan directions.

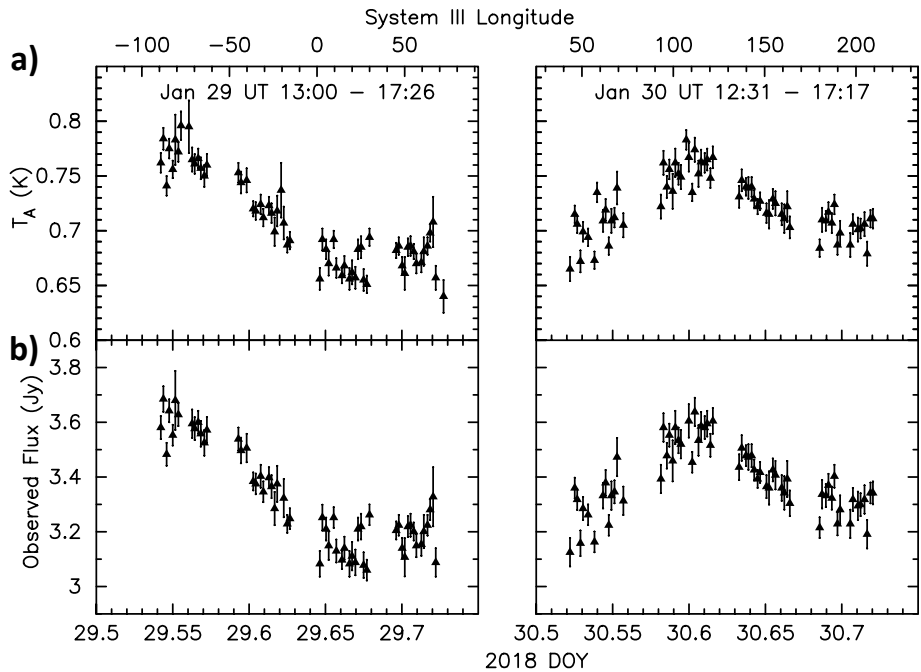

**Figure 3** Examples of the scan antenna temperature, $T_A$(K) and calibrated flux densities (Jy) observed on from January 29 – 30, 2018 when Jupiter was at distance 5.5 AU from Earth. **(a)** The Gauss fit peak antenna temperature. The error bars represent 1-σ measurement uncertainty. **(b)** The scan flux densities (Jy) calculated using the $T_A$ to flux conversion factor listed in Table 1. The error bars correspond to the 1-σ uncertainty for the Gauss fit peak temperatures and the $T_A$ to flux conversion.

## 3 Results and Discussion

### 3.1 Non-thermal Synchrotron Emission

Jupiter's radio emission at wavelengths > 6cm (<5 GHz) is dominated by non-thermal synchrotron radiation from the radiation belts, while at shorter wavelengths dominated by thermal radiation from Jupiter's atmosphere (c.f. de Pater, 1990). Synchrotron radiation is emitted by high energy (MeV) electrons of (energy, E) gyrating around magnetic field lines (strength, B in gauss) at a frequency (in MHz) given by (cf. de Pater 1990),

$$f_p = 4.66\ E^2 B \quad (1)$$

The emission observed in the GAVRT band, at 13 cm, would require $E^2B \sim 500$; in other words, if the magnetic field $\sim 0.5$ G the electron energy needs to be $\sim 30$ MeV. Thus, at regions within the radiation belt where the magnetic field strength is lower, still higher energies are required for the gyrating electrons. For a dipole magnetic field of Jupiter, the field strength decreases as $r^3$ as the planetary distance, r, within the radiation belt increases. Thus, brightness of the synchrotron emission is determined by a combination of field strength and electron energy spectrum; that is, different electron energy populations emit at regions of different magnetic field. Because the 13 cm GAVRT 34-m antenna beam is large (13.6 arcmin compared to < 2 arcmin size of JSR source) the measured flux densities correspond to synchrotron emission from the entire extent of the radiation belt (up to radius $\sim 3R_J$). The synchrotron radiation is highly beamed towards the forward direction of the gyrating relativistic electrons as seen in the sketch of the radiation belts shown in Fig. 4. Furthermore, in Jupiter's magnetosphere most electrons are confined to magnetic equatorial plane and their trajectories depend on their energy, field strength & geometry. Consequently, the strength of the synchrotron flux density as seen from Earth would depend on the orientation of Jupiter's magnetosphere (radiation belts) with respect to the observers' line of sight (see Fig. 4). Jupiter's magnetic field is approximately dipolar, offset from the planet's center by 0.1RJ toward longitude $\sim 140°$ and inclined at 10° from the rotation axis and the surface field strength $\sim 10$ G (c.f. de Pater, 1990). The maximum flux density is observed when the observer direction is in the magnetic equatorial plane. Therefore, we expect the observed flux density to vary with the rotation of Jupiter (9.9 hr period) periodically, passing through maxima and minima. The flux density minimum occurs whenever Jupiter's North or South magnetic pole is pointing towards the direction of the observer.

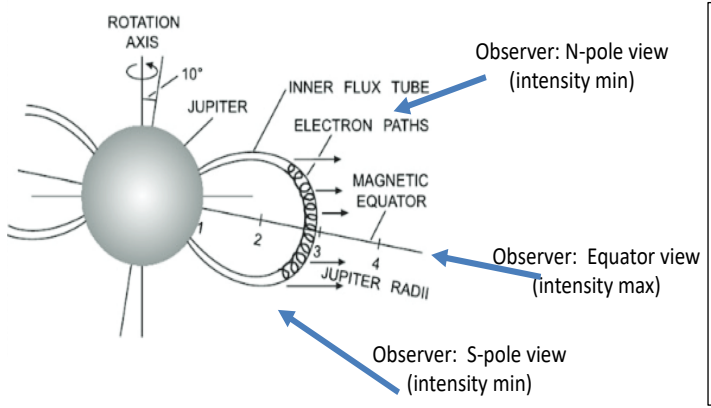

**Figure 4.** Sketch of Jupiter's radiation belts and the synchrotron radiation from electrons gyrating around the magnetic field lines (adapted from Fig. 3.5 in Wilson et al. 2013). As the planet rotates the observer's view also changes with the System III longitude, resulting in corresponding intensity (flux density) variations, with maxima near equatorial view and minima near polar view.

The flux density variation with Jupiter's rotation is referred to as the "beaming curve", in which the observed flux density is described as a function Jupiter's central meridian longitude as seen by the observer or the System III longitude ($\lambda$). Detailed observations of the beaming curves provide a valuable constraint on Jupiter's magnetic field geometry and relativistic electron population in the radiation belts. Here we use our GAVRT data sets to derive the beaming curves at 5 epochs with different Jupiter-observer geometries (see Table 2).

To provide a well sampled data set covering several Jupiter rotations, we use all the GAVRT data observed during the period of each epoch as listed in Table 1. First, we normalized all the flux measurements to a standard distance of 4.04 AU. To extract the non-thermal flux densities (that correspond to the synchrotron emission from Jupiter's radiation belts) we subtracted normalized thermal flux density of 2.02Jy, at 4.04 AU (corresponding to an equivalent temperature of 305 K at 13 cm wavelength (de Pater & Massie, 1985)). The normalized non-thermal flux density plotted as a function of System III longitude (synchrotron beaming data) is shown in Figure 5a. The panels show all, scan by scan, flux measurements along with their $1\sigma$ uncertainty, observed within the period of each epoch. The panels in Fig. 5b show flux densities averaged in 6° wide bins using all scan data and sampled every 6° interval of the System III longitude. Following the approach of Klein et al. (1989) to fit the shape of the beaming curve (flux density as function of System III longitude) we use a Fourier expansion of the form:

$$S(\lambda) = S_0 [1 + \Sigma A_n \sin[n(\lambda+\lambda_n)]] \text{ for n = 1 to 3} \quad (2)$$

where $S(\lambda)$ is the observed non-thermal flux density (normalized to 4.04 AU) for System III longitude, $\lambda$ in degrees; $S_0$ represents a mean value of the synchrotron flux and the amplitude and phase of the flux modulation are described by the Fourier series with coefficients $A_n$ & $\lambda_n$. In Table 2 we list the Fourier coefficients derived by fitting Eq 2 to the observed non-thermal flux data in each longitudinal bin (Fig. 5b). The fits shown in Fig. 5b help us to understand the outliers and the scatter (error bar) in the flux density measurements as discussed below in section 3.2.1. While $S_0$ is the measure of the overall strength (brightness) of JSR, the Fourier coefficients, $A_n$, $\lambda_n$ describe the amplitude and phase of the beaming curve which is also a function Jovicentric Earth declination $D_E$, as marked in the panels in Figs. 5 & 6. $S_0$ is useful to study long term time variability of JSR flux (see section 3.2.1). For a comparison of the results in this paper with those for an earlier epoch observation, in Table 2 we list the coefficients for the 1994 data ( from Klein et al. 1995), also observed with similar $D_E$. The Fourier coefficients are believed to be stable, at least over a 23-year history (Klein et al. 1995).

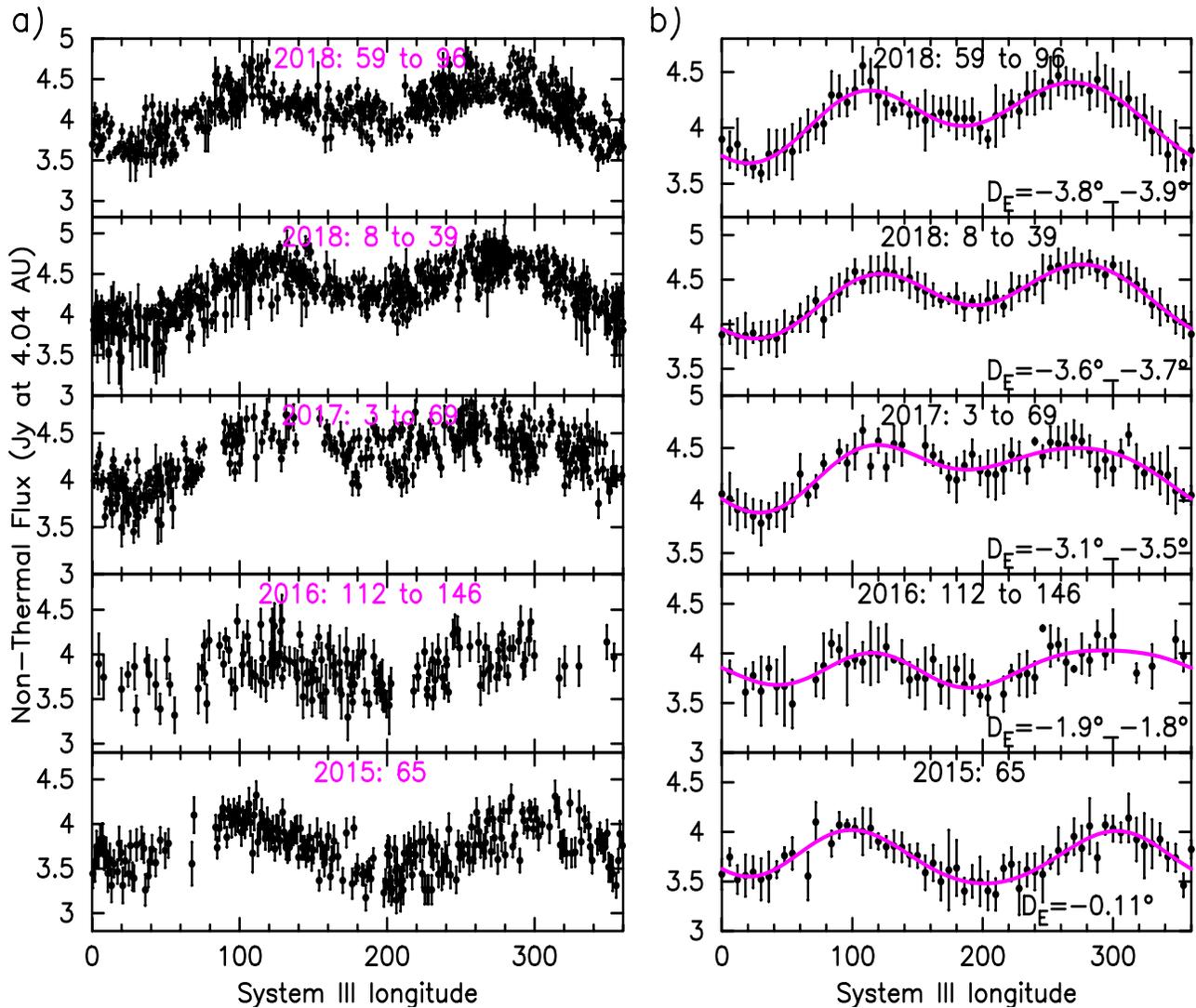

**Figure 5**. **(a)** The synchrotron radiation (non-thermal) flux densities normalized to 4.04 AU. The panels show all measured scan data points observed within each epoch (as labeled) which are plotted along with the 1σ error bar as function of System III longitude.
**(b)** Flux density averaged over 6° wide longitude bins is plotted with the 1σ error bar. The solid line (red) represents a fit to the data using a Fourier series expansion (Eq 2).

### 3.1.1 Modeling Jupiter's synchrotron radiation: Beaming Curves

For a qualitative evaluation of how well the beaming data presented here fit with models for the radiation belts, we compare the beaming flux density data from our observations with those predicted by the models. The basic model developed by Divine & Garrett (1978) for the radiation belt, namely, the production and trapping of charged particles confined by the magnetic field lines, has been verified and updated several times (e.g. Bolton et al. 2001; Garrett et al. 2003) and most recently by Santos-Costa et al. (2017) and Adumitroaie et al. (2016, 2019a). In general, all model computations describe qualitatively the periodic variation of the synchrotron emission with Jupiter's rotation (referred to as the beaming curve) as observed from Earth. However, these computations are being updated incrementally by adding constraints from new data. Furthermore, the beaming curve varies significantly with the declination of the Earth as seen from Jupiter, denoted $D_E$ (Klein et al., 1989; Dulk et al., 1999). Therefore, it is important to compare the beaming data from our observation with model computations for similar values of $D_E$ and distance to Earth. In this paper we compare our beaming data with the model beaming curves generated by us as considered by Adumitroaie et al. (2019a).

**Table 2:** Fourier series fit to observed Jupiter synchrotron radiation (JSR) at 13 cm**

| Parameter | Epoch of GAVRT observation: YYYY_DOY_DOY | | | | | 1994 Feb -Mar[##] |
|---|---|---|---|---|---|---|
| | 2015_65 | 2016_112_146 | 2017_3_69 | 2018_8_39 | 2018_59_96 | |
| Distance (AU) | 4.46 – 4.47 | 4.71 – 5.19 | 4.60 – 5.51 | 5.37 – 5.85 | 4.57 – 5.05 | 4.96-4.60 |
| $D_E$ (deg) | -0.11 | -1.74 – -1.88 | -3.22 – -3.51 | -3.60 – -3.73 | -3.80 – -3.88 | -3.8 |
| $S_0$(Jy) | 3.754 ± 0.093 | 3.861 ± 0.104 | 4.301 ± 0.095 | 4.310 ± 0.003 | 4.105 ± 0.091 | 4.300 |
| $A_1$ | -0.021 ± 0.035 | -0.018 ± 0.039 | 0.045 ± 0.030 | 0.050 ± 0.001 | 0.046 ± 0.032 | 0.046 |
| $A_2$ | 0.064 ± 0.034 | 0.047± 0.036 | 0.046 ± 0.030 | 0.065 ± 0.001 | 0.060 ± 0.032 | 0.049 |
| $A_3$ | -0.012 ± 0.035 | -0.015 ± 0.037 | -0.010 ± 0.030 | 0.006 ± 0.001 | 0.007 ± 0.032 | 0.009 |
| $\lambda_1$ (deg) | 245 ± 95 | 326± 118 | 243 ± 40 | 231 ± 13 | 239 ± 39 | 237.98 |
| $\lambda_2$(deg) | -66 ± 16 | -70 ± 27 | -67 ± 20 | -65 ± 1 | -58 ± 15 | -66.86 |
| $\lambda_3$(deg) | 9 ± 54 | -86 ± 49 | -15 ± 61 | 24 ± 3 | 280 ± 90 | 8.55 |

**Note: Coefficients were derived from curve fit solutions to ensemble of data collected over several days as indicated:(S= $S_0[1 +\Sigma A_n \sin[n(\lambda+\lambda_n)]]$, for n = 1 to 3, where $\lambda$ is the System III longitude)

[##] Data from Klein et al. (1995), coefficients re-calculated to match Eq 2

As illustrated in Fig. 4 Jupiter's synchrotron radiation environment is produced by a complex physical system, which is not yet completely understood. Modeling of the radiation belt is needed to understand the beaming data observed from Earth (for example as shown in Fig. 5a & b), as well as the intensity and polarization maps observed for example, with the Very Large Array (VLA) by de Pater & Jaffe (1984). From Juno perspective such modeling is required not only to understand the charged particle environment, but also to separate synchrotron emission from MWR's observations for thermal emission of the planet. Modeling JSR beaming requires multiple refinements, based on the *in-situ* data, of a higher fidelity model for the synchrotron emission, namely the multi-parameter, multi-zonal model, for example as in Levin at al. (2001). Prior to the Juno mission such models relied on an empirical electron energy distribution, which was constrained by VLA observations. Our modeling effort includes tuning the multi-zonal parametrization adopting an inverse modeling approach given multiple orbit measurements of the synchrotron emissions from the Jovian radiation belt and the planet's magnetic field. This effort makes use of the cumulative perijove observations to optimize the electron distribution parameters with the goal of matching the synchrotron emission observed along Juno's Microwave Radiometer's (MWR's) lines of sight. Currently this modeling is still ongoing and will need to be regularly updated as additional observations from the Juno MWR and magnetometer instruments become available.

Adumitroaie et al. (2016, 2019a) presented the framework for an extension of the Levin et al. (2001) multi-zonal, multi-parameter model from Earth-based observers to Juno spacecraft point of view, and to simulate the four Stokes parameters of the synchrotron emission using assumed electron distributions and Jovian magnetic field models (originally, VIP4 and now the latest version, JRM09). This model incorporates magnetic-field-derived quantities such as the M-shell and $B_{crit}$. M-shell is a nomenclature update on the classic terminology of L-shell to reflect the consensus in the literature when referring to equatorial crossing distances in non-dipolar magnetic fields. The parameter $B_{crit}$ is defined as the minimum magnetic field amplitude for a given M-shell intersecting the upper boundary of the atmosphere. As such, $B_{crit}$ determines the critical pitch angle representing particle loss/scattering, i.e. particles with a pitch angle less than critical pitch angle or more than the supplement of critical pitch angle are assumed lost when they mirror at or below the upper boundary of the atmosphere. For the JSR predictions we used the set of the ancillary parameters (M-shell and $B_{crit}$) computed in the framework of JRM09 magnetic field model (Adumitroaie et al, 2019a). It may be noted that L-shell in Adumitroaie 2019a is retitled M-shell here according to a recent convention in the literature. Following this approach, we calculated the JSR beaming curves for each epoch for an Earth-based observer declination, $D_E$ and distances as listed in Table 2.

In Figure 6. we show the observed beaming data along with the overlay of the model beaming curves. The model curves seem to fit the observations well. To further quantify the model fit to the data, in Fig. 6 we also show the linear Pearson correlation coefficient, r, between the data and model curve. The lowest value r = 0.58 for the epoch 2016:112_146 is due to the poor scan observations for the longitudes 300° to 70° caused by bad observing conditions during the first 1 hr and 3 hr on DOY 112 and 146, respectively. Nevertheless, for this epoch there is good agreement between data and model for the well observed longitudes 90° to 300°. Overall the consistency between the observed beaming data and the model curves over 5 epochs and at different $D_E$ validates the JSR modeling by Adumitroaie et al. (2019a).

To examine the improvements achieved in our current model beaming curves, in Figure 7 we compare our results for one of the epochs, 2018:008_096 with beaming curves derived by Levin et al. (2001) for $D_E$ = 3.8°. There exist numerous model computations developed over the years, but here we use Levin 2001 as a proxy for earlier computations. The beaming curves derived by Levin2001 and by us (this paper) are over-plotted on the observed beaming data. Clearly, our current model curve shows remarkable improvement over Levin2001 and provides better match to the observed beaming data. The double peaks and overall profile of our beaming curve is broadly consistent with the model. The apparent symmetry of the double peaks as in the model curves suggests no need for longitudinal asymmetry in the particle distributions. Our 2018 data show much shallower minimum (at longitudes near 200°) than in the model prediction. To further validate our current model beaming curve in Fig. 7 we show the beaming data observed in 1994 (Klein et al. 1995), also observed for $D_E$ = 3.8°. Note that we didn't have access to 1994 data and therefore, we reconstructed the beaming curve using the Fourier coefficients given in Klein et al. paper. Indeed, the 1994 data (blue line) shows good correspondence to our model curve (red). Overall, our beaming data and model curves show the new magnetic field model JRM09, as an integral part of the synchrotron radiation prediction. Future GAVRT beaming data combined with Juno MWR and other ground-based observations with LOFAR and VLA will be used as improved model constraints and model validation for the JSR electron belt.

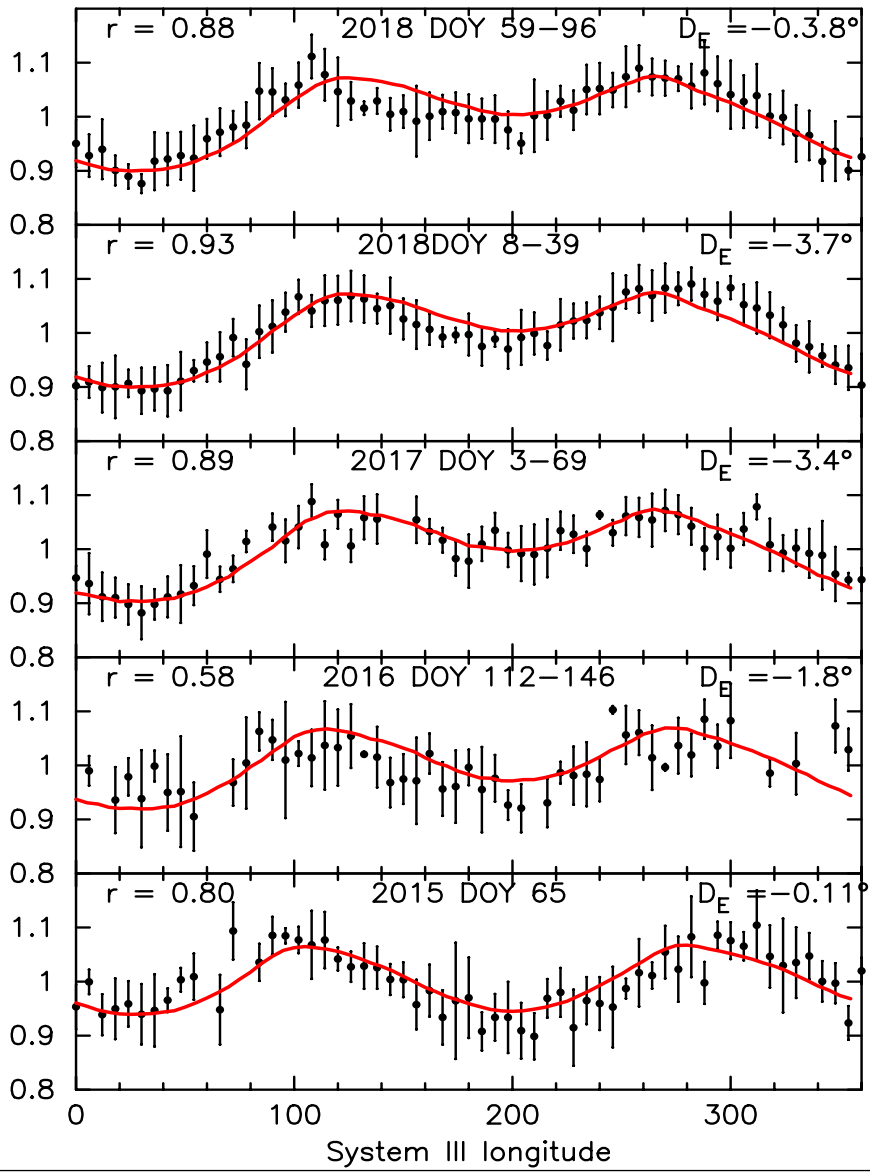

Fig. 6 In each panel; the observed data in longitude bins (shown in black) at $D_E$ as labelled are compared with corresponding model beaming curve (shown in red) derived incorporating some new information from Juno data (Adumitroaie et al. 2019a). The linear Pearson *correlation* coefficient between data and model, r is listed on top left of each panel. The data seem to correlate well with the model beaming curves.

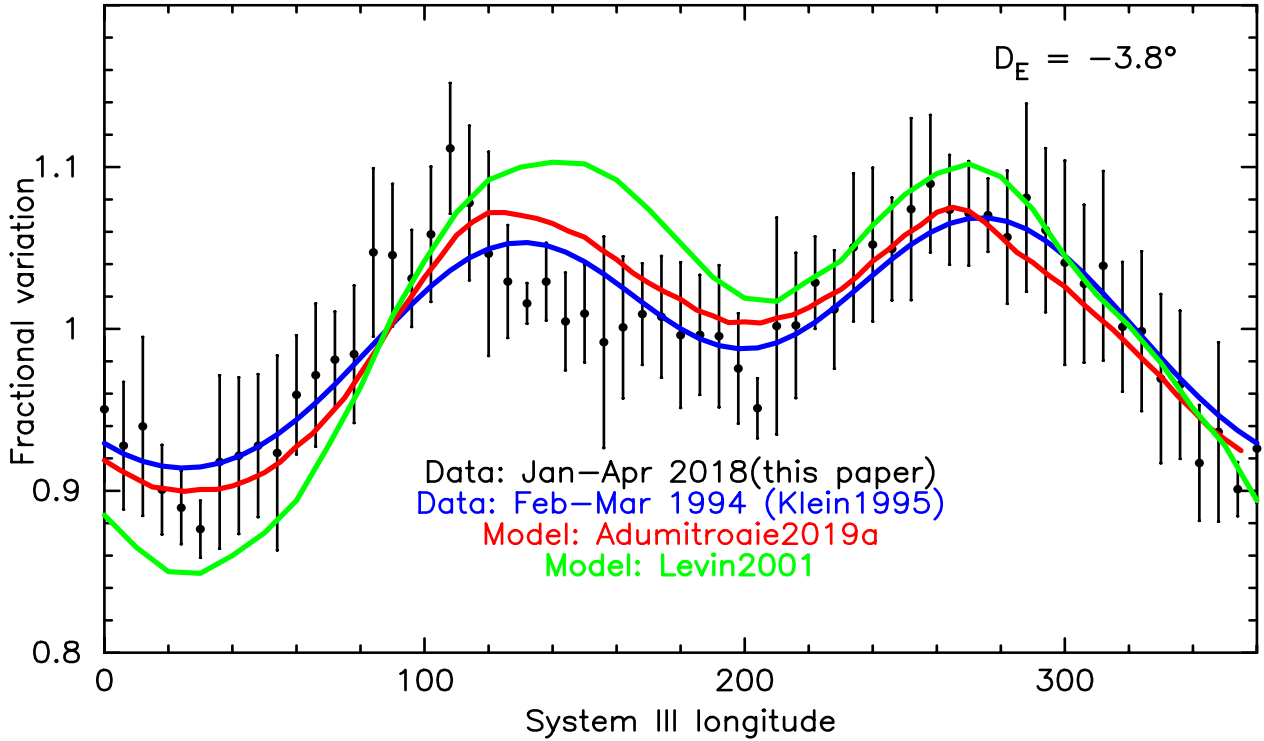

**Fig. 7** Comparison of data with synchrotron model beaming curves. The observed (data) beaming curves for two epochs: Jan -Apr 2018 (this paper shown in black) and Feb-Mar 1994 (from Klein et al. 1995, shown in blue). At both epochs the declination of Earth, $D_E$ was -3.8°. An earlier model curve derived by Levin et al (2001), shown in green. The recent model curve (Adumitroaie et al. 2019a) based on the JRM09 magnetic field model (Connerney at al., 2018) updated from Juno data is shown in red. Both model curves were derived for $D_E$ = -3.8°.

### 3.2 Monitoring JSR flux densities: Long- & short-term variations

Since the confirmation of JSR flux density variability in 1972 (Klein et al. 1972), NASA DSN initiated a long-term observational program to monitor the flux density (de Pater &Klein, 1989; Klein, Thompson & Bolton, 1989). These results paved the path for investigating plausible correlation between JSR and solar activity. It may be noted that indeed, the GAVRT observations presented here are part of a reincarnation of this DSN monitoring after GAVRT was formed. The early studies of the variability by Bolton et al. (1989) indicate that there exists a time lag from months to years between radio measurements and solar wind parameters, especially the ram pressure showing the strongest correlation with a time lag of a few years. To what extent Jupiter's magnetosphere is influenced by the solar wind still remains an outstanding question (e.g., Bagenal et al., 2017). Observations from the Juno mission provide a new opportunity to study solar-wind-magnetosphere interactions in situ at Jupiter. For example, the Juno spacecraft crossed Jupiter's bow shock and into the magnetosheath on 24 June 2016. The 53-day polar orbits give us a unique opportunity to survey Jupiter's dawn magnetosheath during the perijove crossings (Ranquist et al. 2019). Studying the magnetospheric processes with direct evidence of solar wind influence, for example, the motion of Jupiter's bow shock and magnetopause, the opening and closing of magnetic flux in the outer magnetosphere is important to understand the features of a solar wind-driven magnetosphere. Furthermore, including a significant dawn–dusk asymmetry as observed is relevant for the time evolution (history) of the JSR (e.g. Han et al. 2018). Here we use our multi epoch observations between 2015 and 2018 to examine detectability of plausible variability in the JSR flux density and to plan continuation of the GAVRT monitoring program during the duration of the Juno mission.

### 3.2.1 Flux residue after removing beaming curve & measuring daily flux density:

One major difficulty in monitoring the JSR flux density on a daily basis is that it requires a number of measurements over periods of more than 9.9 hr (Jupiter's rotation period), which is seldom feasible with a single GAVRT session. For a reliable estimate of the daily flux we need to sample, each day, the full beaming curve covering longitude range 0°- 360°. For their variability study, Klein et al. (1989) use the peak values in the beaming data normalized to distance 4.04U. However, in the present GAVRT data, except for DOY 65 in 2015, only a few flux measurements are available on each day. Therefore, it is necessary to combine several days of measurements to compile the beaming data as presented in Fig. 5a. To estimate a daily average value of the flux for the purpose of looking for variability, we adopt the following procedure: First we treat the $S_0$ from the fit to the beaming data at each epoch (Fig. 5b & Table 2) as the mean flux for the entire duration of the epoch. We then can compute the residual flux in the individual scan-by-scan measurement by subtracting the mean beaming curve from observed fluxes (in Fig. 5a) by matching the System III longitudes for each scan. To illustrate this approach, in Fig. 8a we show a plot of non-thermal flux density measured by each scan observed on DOY 29-30 in 2018 along with the corresponding mean beaming curve for Epoch 2018_008_039. The fits to the beaming data in Figure 5a show several data points with large deviation from the fit. It is important to understand these outliers as they could be caused by intrinsic variations in Jupiter's flux density, fluctuations in the background sky contribution, and/or errors in the flux measurements (scans). Example of the residual fluxes obtained after subtracting the fitted mean beaming curve are also shown in Fig. 8a. Note that the observed data do not fully cover the full longitude range. Nevertheless, the mean beaming curve for DOY 8 – 39 seems to fit the profile shape well, but indicates an "excess" flux ~ 0.15 Jy over the mean beaming curve (see Fig. 8c). This non-zero flux residue would suggest detection of an increase on these two days above the epoch mean flux density, $S_0$. Thus, we can estimate the JSR flux density for a given day by adding the average flux residue for the day to the epoch fit to $S_0$.

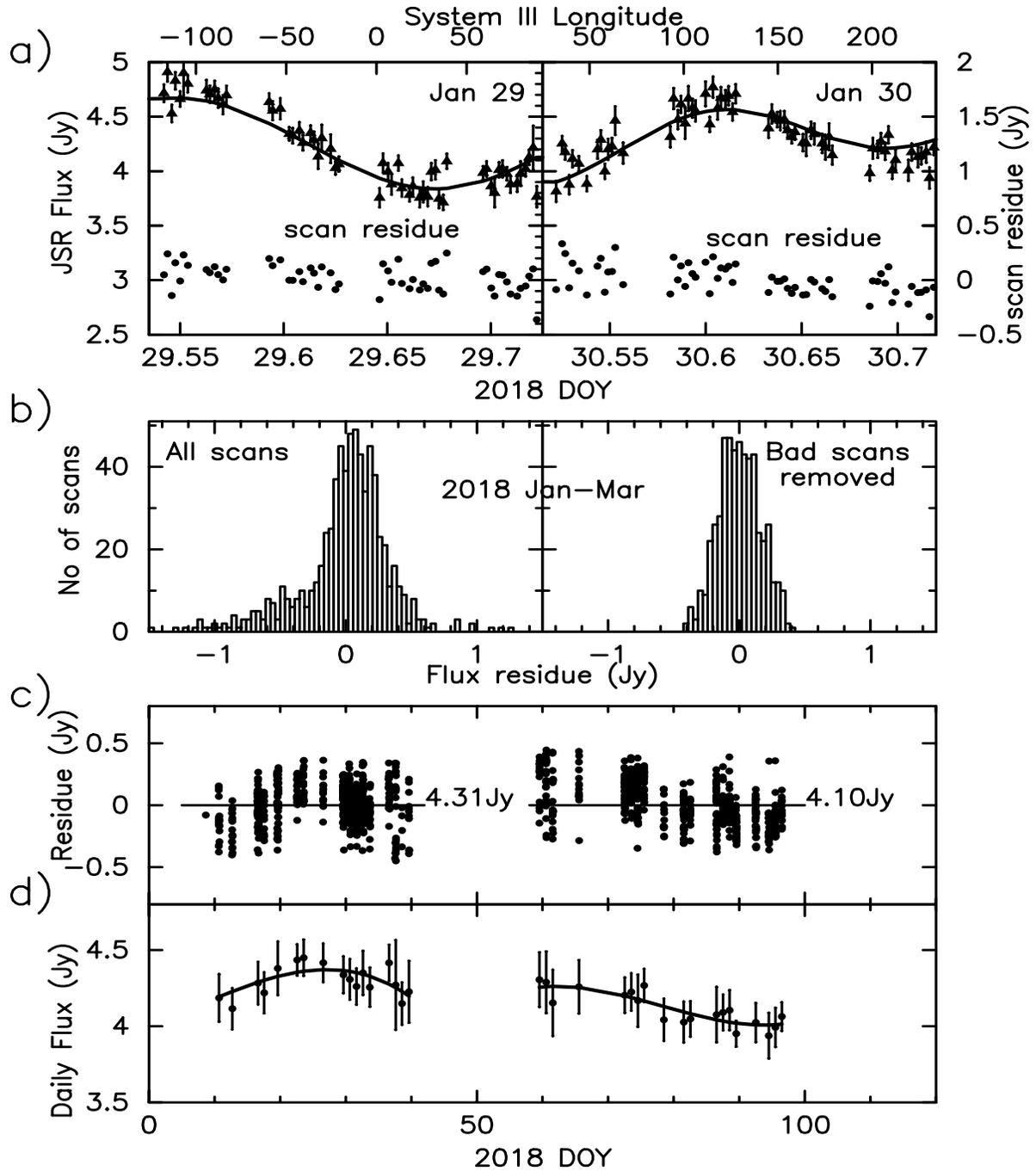

**Figure 8** **(a)** Non-thermal flux observed by each scan and the corresponding residual flux plotted for DOY 29-30. The solid lines represent the mean beaming curve as shown in Fig. 5b. **(b)** Histogram of the residual flux in each scan for the epoch observation 2018_008_039: *(left panel)* all observed scans showing a distribution skewed towards low measured flux density; *(right panel)* data used for final analysis after all bad scans were removed. **(c)** Daily scan by scan residual flux. **(d)** Estimated daily flux density plotted for all observations in 2018. The solid line is a fit to daily averages, suggesting short term variations

A more rigorous method would be to use the beaming curve fit to the observations of each individual day to determine $S_0$. However, for many of our observing sessions, the duration and longitude coverage are not sufficient to fit a beaming curve. Removing an epoch-wise mean beaming curve from each daily session is algebraically equivalent to fitting a daily beaming curve, assuming that the shape of the curve is approximately constant over the epoch.

In addition to measuring trends within an observing epoch, computing the flux residue as described above is also useful in evaluating individual scans within an observing session. In the scan by scan residual flux data, the outliers in the flux measurement stand out more clearly. For example, a histogram distribution of the flux residue for all scans observed (left panel in Fig. 8b) shows a peak somewhat broader than for an overall 1-$\sigma$ measurement uncertainty of 0.02 K or ~0.18 Jy in the normalized flux densities. However, there is some asymmetry in this distribution, between the lower and higher sides of the peak flux residue. Such skewed distribution with more cases of negative values over positive is clear evidence that some of the scans measured low flux values, possibly due to poor pointing. While a pointing calibration is done at the start of each observing session, pointing can still be degraded after switching between sources and after an extended period of tracking. However, the measurements with excess positive residual could be significant as intrinsic variations in Jupiter's flux density or fluctuations in the background sky contribution. The scans thus identified with extreme pointing issues, by the low flux and/or large offset in Gauss fit peak, though few, were removed from the analysis. The right panel in Fig, 8b shows the histogram of flux residue analyzed using the data after removing all identified bad scans. The overall distribution in consistent with the daily scan flux residue plotted in Fig. 8c. We did not notice any such observation issues with 3C353 which had stable $T_A$ (flux density) measurements, as seen in Figure 2. Because of Jupiter's lower flux than calibration sources (3C353), we may need a better pointing strategy for future Jupiter scans.

Figure 8c & d further illustrates how we estimate the daily measurement of JSR flux densities from a limited number of scans observed on each day. In Fig. 8c the scan by scan flux residue (black dots) observed during the two epochs in 2018 are plotted against the DOY. The horizontal line (zero residue) corresponds to the mean epoch flux density $S_0$. Adding the mean flux residue to $S_0$ for each day of observation we are able to estimate the daily JSR flux densities. The day flux densities along with the 1-$\sigma$ scatter in the daily scan data are shown in the lower panel in Fig. 8d. The polynomial fits to the day flux density (the solid lines) to these epoch data seem to indicate variations at the level of ~ 10% on the scale of 10-20 days. This variation is well above the confusion limit and lasts over several days, so it is unlikely to be contamination from point sources which shows day-to-day variation.

No correction for antenna beam effects was applied to the measured flux densities. The 34-m S-band beam (13.6 arcmin) is large enough (HPBW/$R_J$ > 36 at the shortest distance to Jupiter in the present observations) to ignore any source size correction (cf. Table 1 & Fig. 1 in de Pater et al. 2003). Flux measurements could also be contaminated by fluctuations in the sky background and sources in the main beam or side-lobes near Jupiter (de Pater & Butler, 2003). However, such contribution should be reduced in our data because of using scanning mode observations instead of nominal On-Off measurements. For example, the sky background is well subtracted out by the baseline fitted to scan data. However, any weak point source within the beam at Jupiter's sky position on a given day will contribute to the measured flux density. Such a contamination will vary on daily timescales because Jupiter's motion with respect to sidereal sources is ~ 1 beam size per day. This only adds to the uncertainty in the day to day flux measurement. The contribution from such background confusion was determined by Klein et al., 1972, to be 0.15 Jy for the 20 arcmin beam of the DSN 26 m antenna. Scaling this to the smaller 13.6 arcmin beam of GAVRT 34 m yields an estimated confusion error of 0.07 Jy. Thus, the confusion signal is typically at the level < 2.5% of Jupiter flux (~ 4 to 6 Jy) for the GAVRT telescope (c.f. Klein et al.1989). Because of the random nature of presence of background sources this will vary from day to day, with the occasional stronger background source. To quantify further any background point source contribution to the measured daily JSR flux density, we searched the 1.4 GHz NRAO VLA Sky Survey (NVSS), which is available on line (Condon et al. 1998), for point sources with flux > 10 mJy at the position of Jupiter on a given day. The estimates of background point source contribution are plotted in Fig. 9c. The background sky contributions seem negligibly small on most days, the highest being 0.12Jy and will be even smaller when extrapolated to 2.28 GHz. Thus, the daily flux variation persisting over a few days as in Fig. 8c & 9b is not due to background sources and is significant as source variability than an observational issue. Ideally, it is desirable to quantify any sky contribution by (i) simulating GAVRT beam on available sky maps and, (ii) re- observing Jupiter's sky position on a given day after Jupiter has moved out. However, it is beyond the scope of this paper and such an approach is more appropriate for future analysis of variability in the time frame of Juno perijoves.

### 3.2.2 JSR time variability and solar wind interaction:

The timescales of various synchrotron emission processes, including the radial transport, are quite vast in range. Monitoring of long-term variations of JSR flux density is the key to understand its correlation with solar activity and solar wind in particular (c.f. Bolton et al. 1989). To investigate the variability in our data observed at multiple epochs from 2015 to 2018, we measured the daily mean flux density for all observation days, using the fitted beaming curves and the flux residuals, as described above. In Figure 9b are shown the day averages of the JSR flux density measurements derived from GAVRT data for the period 2015 to 2018. Our results show a clear trend in JSR flux density increase from March 2015 to April 2018.

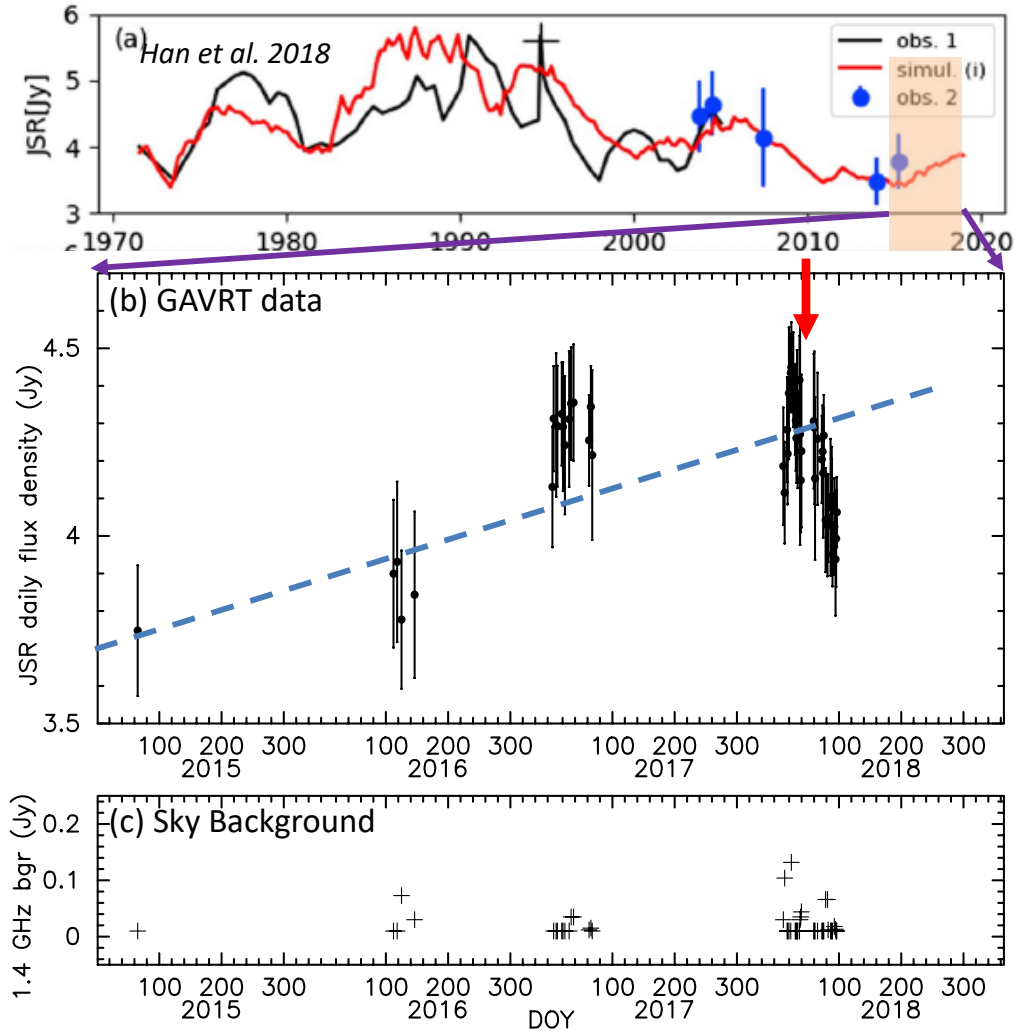

**Figure. 9 (a)** Simulated JSR (red) and observed flux density (black line and blue dots) as a function of time (reproduced from Fig. 3a of Han et al. 2018, and see the references therein for the data). The shaded area marks period overlapping GAVRT monitoring shown in the lower panel. **(b)** GAVRT monitoring of Jupiter's synchrotron radiation (JSR) during 2015 to 2018. Day averages of the non-thermal flux density (see text) are plotted as a function of DOY for each year. The red arrow shows examples of short terms variations. Note the JSR increasing trends in the data in both the simulation and GAVRT data. **(c)** Estimate of plausible background point source contribution to the measured daily JSR flux densities (see text). It is negligibly small on most days, with a maximum of 0.12Jy .

We interpret this result as evidence for plausible changes in the electron population within the radiation belt induced by solar wind interaction. The long-term monitoring of EUV from Io plasma torus by the spectrometer EXCEED (EXtreme ultraviolet spectrosCope for ExosphEric Dynamics) on board the HISAKI satellite (Yamazaki et al., 2014; Yoshikawa et al., 2014) led to understanding that the changes in solar wind ram pressure drive fluctuations of dawn-to-dusk electric field in the Jovian magnetosphere (Murakami et al., 2016). Recently, Han et al. (2018)

have investigated the temporal variations of JSR modeling the influence of fluctuating dawn-to-dusk electric field on the radiation belt. Furthermore, they use their diffusion model to simulate JSR flux densities for the period from 1971 to early 2018. Fig.9a shows one of their simulations, for radial diffusion coefficient of 3 x $10^{-10}$ $L^3$ $s^{-1}$ where L is shell parameter, < 6 (see Han et al. 2018). They also find a correlation coefficient of 0.6 between the simulated JSR and the ground observation data prior to 2005. As seen from this plot the data available after 2005 is sparse and our present results will be critical to future studies. Fortunately, Han et al. simulations include the predictions of a trend in JSR variation for the years 2015 to 2018 (see the hatched area in Fig. 9). Interestingly, the observed trend, in JSR flux density in our GAVRT data, is consistent with this simulation for solar wind interactions. Thus, it seems likely that the dawn-to-dusk electric field associated with solar wind conditions drives long- term variations of JSR.

In Figure 9b (feature highlighted by the downward red arrow) we identify a short-term flux density variation in the GAVRT data for epoch 2018 DOY 8 to 96. Though Han et al. (2018) simulations extend up to early 2018, the plots do not show this short duration variation. This is not surprising because their assumed value for the radial diffusion coefficient precludes simulating variability on such scale. The long-time scales for radial diffusion, of the order of years, make it is easier to explain JSR long-term variations. However, the short-term changes in JSR likely result from intrinsic changes in energetic electron population induced by physical processes, for example, the solar UV/EUV heating of Jupiter's upper atmosphere. In Tsuchiya et al. (2011) model the short-term variations can be explained in terms of an enhanced diffusion coefficient, taking account the observed enhanced solar UV/EUV flux. Short-term variations in JSR (~days) have been observed in response to solar EUV flux changes (c.f. Kita et al. 2013; Bhardwaj et al. 2009). However, recent GMRT (JSR)/EXCEED (UV/EUV) results indicate that variations, on time scales of weeks, show no obvious correlation to either EUV or dawn-dusk electric field changes (Kita et al. 2019).

The short-term variation in our GAVRT data (Fig. 9b) seem significant enough for further investigation. The F10.7 radio flux is regarded as good proxy for solar UV/EUV flux. Therefore, we examined the 10.7 cm flux density data (available at https://www.spaceweather.gc.ca/solarflux/sx-5-en.php) looking for any correlation with the JSR flux variation observed in GAVRT data. No correlation was found with a few weeks lag, though there was a weak event with a month lag time. Contiguous JSR monitoring on a longer time baseline would be need to draw any useful conclusion. Interestingly, the GAVRT epoch observations (2018 DOY 8 -96) shown in Fig. 9b includes the perijove (PJ11), 2018 DOY 30-38. During this perijove the Juno instruments detected a few-day (at $R_J$ ~ 30 to 60) transient (enhancement) of both the azimuthal field and radial magnetic field, as well as plasma density and temperature (Nichols et al. (2020). It is suggested that this enhancement could have been possibly induced by a solar wind compression. Of course, it is not at all evident that such transient enhancements observed by Juno instruments will immediately manifest in JSR flux changes. Nevertheless, our results clearly demonstrate how GAVRT monitoring program is a useful resource to study variability in Jupiter's synchrotron emission, and to check and validate the Juno results on electron distribution and magnetic field models, as seen from a far (e.g. Earth) vantage point.

## 4  Conclusions

In this paper we presented a frame work for analyzing the data observed under the GAVRT program designed to monitor Jupiter with teacher-student participation. Our results demonstrate the usefulness of the data taken by GAVRT for frontier science research, especially in the context of the Juno mission. We found that the consistency between the observed beaming data and our model beaming curves over 5 epochs and at different $D_E$ validates the JSR modeling by Adumitroaie et al. (2019a) and the use of magnetic field model JRM09. Here, we presented results using observations made at irregular intervals of time and with few scans on a given day. We derived daily mean JSR flux density using only a few scans observed on a given day and removing the beaming effect using mean beaming data observed over the period of an extended epoch. The time variabilities reported here, and their interpretation in terms of solar wind interaction are of particular interest to Juno's perspective. The extent of the solar wind's influence and how it may drive internal processes in Jupiter's magnetosphere is not well understood. However, times of elevated solar activity is expected to show correlations with the auroral radio, optical and X-ray emissions, as well as synchrotron radiation from the radiation belts with some time delays. GAVRT is identified as one of the key resources (e.g. Han et al. 2018) for monitoring JSR and it further emphasizes the importance to plan future dedicated observations, especially during Juno's perijove events and the increasing solar activity in the coming years. It is highly desirable to have an observing capability for faster, uniform sampling and full coverage of System III longitude to provide JSR data sets useful to the measurements of the magnetosphere by the Juno instruments.

**Acknowledgement:** We thank an anonymous referee for critical comments and suggestions that improved the discussion significantly. We acknowledge the support of Lisa Lamb, President/CEO and LCER staff for all their infrastructure & observational support to data presented in this paper. Peyton Robertson spent the summer of 2019 analyzing the 2018 GAVRT data. We thank Dr Fran Bagenal and Dr. Joseph Lazio for critical comments. This work was performed at the Jet Propulsion Laboratory, California Institute of Technology, under contract with the National Aeronautics and Space Administration.*© 2020. All rights reserved.*22